# Improving the quality of active millimeter wave standoff imaging by incorporating the cross-polarized scattering

**SEYEDEHZAHRA SHOJAEIAN, MEHDI AHMADI-BOROUJENI, SHIVA HAJITABARMARZNAKI**

*Department of Electrical Engineering, Sharif University of Technology, Tehran, Iran.*
*\*ahmadi@sharif.edu*

**Abstract:** In this paper, we report a polarimetry-based THz imaging technique that highly benefits from backscatter data reflected from PEC hidden objects, considering the edge diffractions of the object as a key point. First, based on physical optic principles, we investigate the cross-polarized reflections of PEC objects in order to show the superiority of reflections of the edges to other parts of the object. Next, we represent the results of a near filed simulation using Feko to study the differences between cross and co-polarized reflections of a PEC object and the human body beneath it. To a further illustration, our experimental results are presented in which we utilized a THz imaging system consisting of a semiconductor-based THz camera and a horn antenna operating at 100 GHz as the source. Cross-polarized backscatter data is analyzed to distinguish PEC objects from human body. Besides, we used two orthogonal linear polarizations of the source to maximize backscattering from orthogonal edges. Experimental results show a noticeable improvement in edge detection of hidden objects indicating that this method can result in a high accuracy in THz real-time imaging.

## 1. Introduction

THz imaging has been increasingly considered as an attractive research area in recent years due to great demands for imaging applications in security detection, through package detection, pharmaceutical and biomedical fields. Among different frequency ranges, THz range has been attracted a great attention because of its unique features. Not ionizing human cells, THz waves won't cause damage to individuals so it can be used in public places and airports for security purposes [1] or in biomedical imaging for diagnosing different diseases [2]. Also it has a good penetration into nonmetallic materials that makes it possible to detect objects hidden beneath different types of clothing [3] and detecting hidden objects inside different things [4]. As a result, many research groups have been working on improving the accuracy and speed of THz imaging by employing various theoretical and practical methods. Advances in THz biomedical imaging have been occurred by development of Quantum Cascade Lasers (QCL) that relying on its relatively high optical power, enables researchers to investigate various methods to improve imaging quality [5].

THz imaging is being performed using two different techniques, passive THz imaging and active THz imaging. In passive THz imaging, there is no source to illuminate the object and the self-radiation of the object and also reflections from surrounding illuminations like sky are being measured to carry out imaging process. In contrary, active imaging uses a THz source for illumination and measures the backscatter responses. As the passive imaging technique relies on self-radiation of the object, it changes with object's temperature. Also environmental circumstances may have influences on imaging results. Considering these factors , active imaging seems to be a good choice for security applications like stand-off imaging due to its high dynamic range and signal to noise ratio (SNR) [6,7]. Many investigations have been done recently in this area that radar imaging is a common imaging method that has being used by different imaging groups. Jet Propulsion Laboratory (JPL), has made wide progresses in developing high resolution stand-off imaging. They have launched a well-stablished (FMCW) radar system and accomplished some experiments at different frequencies from 340GHz to 675GHz. in [8] they represented a coherent illuminating radar at 580GHz with a phase-sensitive detector and obtained centimeter-scale resolution. In [9], they investigated the effect of focal depth in imaging quality. In [10] they developed a 675GHz system with a new optical design that brought them the possibility of single-pixel imaging and real-time frame-rate. They also made a comparison between imaging features at frequency of 340GHz and 680GHz by stablishing a fully polarimetric system [11].

Multi-polarimetric imaging is one of the most desired techniques in MMW detection for different applications. Several dielectric disks has been studied in [12], to investigate the effect of roughness on co-polarized and cross-polarized backscatter data. Also in [13], the backscatter response of a multi-polarized radar on different outdoor surfaces has been examined, in order to be applied in remote sensing projects. In [14], different backscatter data acquired from co-polarized and cx-polarized reflections is being used for indoor navigation and mapping. Finally as an imaging application, [15] and [16] have realized the effect of cx-polarized reflection on improvement of concealed weapon detection at 340 GHz and 25-30 GHz respectively and [17] performed the same investigation using circular polarized sources operating at 10-20 GHz.

Specular reflection from smooth surfaces like human body is one of the most critical limitations in active stand-off imaging. Polarimetric analysis of the backscattered wave from the object could be a solution to overcome this limitation. In this paper, a polarimetric technique is represented based on utilizing extra information that could be obtained from cx(cross)-polarized reflections of PEC objects. As it is expected that edges of a PEC object will have significantly higher cx-polarized reflections comparing to its flat parts, we investigated this fact in two ways: first, we used the higher intensity of edges in cx-polarized reflection in order to more accurately distinguish the shape of the objects by eliminating the specular reflection from human body. In next step, we compared different linear polarizations of the source to study the possible relation between the source's polarization and the backscatter reflections from different edges of the PEC object. Our imaging system is a quasi-optical stand-off imaging system that has been presented by Terasense Inc. [18]. In our experiments, this imaging system including a 100 GHz source, a THz camera coherently collecting data from the whole target and the lab computer analyzing the received data, along with polarization analysis, are being used to perform the imaging process.

The content of this paper is organized as follows: section 2 gives an analysis of polarimetric idea behind our experiments and its implementation, section 3 represents the near-field simulation results, section 4 illustrates the experimental results and section 5 shows the final result after edge detection and finally section 6 represents our conclusions.

## 2. POLARIZATION ANALYSIS

Polarimetric measurements of backscatter data from different targets show that additional information can be received from analyzing the cx-polarized reflections. As the PEC objects have significant cx-polarized reflections from their edges comparing to the surface of human body, the cx-polarized data can help us distinguish the hidden object from the body particularly when doing it by comparing their co-polarized reflections becomes difficult. Here we use the closed form of physical theory of diffraction (PTD) method presented in [19], to analyze the backscatter response of a PEC object. According to these closed form equations, the diffracted field caused by the edges of the target can be written as:

$$\vec{E}^{PTD} = jk \int_c \left[ \eta(r \times r \times \hat{t}) I_e(\vec{r}') + (r \times \hat{t}) I_m(\vec{r}') \right] \frac{e^{-jk(\vec{r}-\vec{r}')}}{4\pi|\vec{r}-\vec{r}'|} dl' \quad (1)$$

$$I_e = \frac{2\vec{E}^i \hat{t}(D_\parallel - D_\parallel')}{jk\eta \sin^2 \beta'} + \frac{2\vec{H}^i \hat{t}(D_x - D_x')}{jk \sin \beta'} \quad (2)$$

$$I_m = \frac{2j\eta \vec{H}^i \hat{t}(D_\perp - D_\perp')}{k \sin \beta' \sin \beta} \quad (3)$$

Where $D_\parallel$, $D_\parallel'$, $D_x$, $D_x'$, $D_\perp$ and $D_\perp'$ are diffraction coefficients represented in [20] and $I_e$ and $I_m$ are the electric and magnetic currents initiated by diffracted electric and magnetic fields on the edge. Diffracted fields and β can be seen in Fig. 1

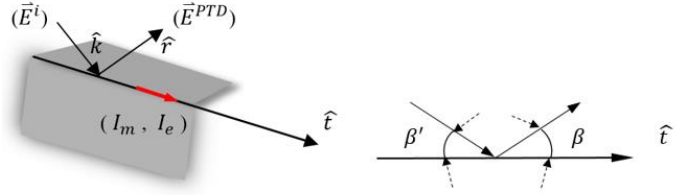

Fig. 1. Interpretation of components used in PTD equations

By implementing x-polarized and y-polarized waves as $\vec{E}^i$ in equation (1) and deriving diffracted fields for both horizontal and vertical edges, it can be found that for each polarization of the source, the edge with the same direction with the polarization of the source will scatter cx-polarized reflection more than the opposite edge. In the next part, we presented the Matlab code implementation of the imaging process using Eq. (1) to demonstrate our claim. In this implementation, we illuminated a 5×5 cm PEC square and a 5cm diameter PEC disk by a 100GHz x-polarized plane wave with the illumination angle θ = 30° relative to z-axis. Then we calculated the cx-polarized(y-polarized) reflections of both horizontal and vertical edges of the objects on the x-y plane at $z = 2$ m for the PEC square and $z = 3$ m for the PEC disk. Next, we passed the calculated waves through the collimator lens by adding the quadratic phase of the lens to it. Distribution of transmitted wave behind the collimator lens can be written as [21]:

$$U_l'(x,y) = U_l(x,y) \exp\left(-j\frac{k}{2f}(x^2 + y^2)\right) \quad (4)$$

Where $U_l(x,y)$, $U_l'(x,y)$ and $f$ stand for calculated E-field before the lens, E-field just after passing through the lens, and focal point of the collimator lens respectively.

As the final step, using the Fresnel integral, we formed the final images at $z = 35$ cm behind the lens. Combining Eq. (4) and Fresnel integral, one can extract the equation to derive the final image [21]:

$$U_f(u,v) = \frac{\exp\left[j\frac{k}{2z}(u^2+v^2)\right]}{j\lambda z} \int\int_{-\infty}^{\infty} U_l(x,y) \exp\left(j\frac{k}{2}(\frac{1}{z}-\frac{1}{f})(x^2+y^2)\right)$$

$$\times \exp\left[-j\frac{2\pi}{\lambda z}(xu+yv)\right]dxdy \quad (5)$$

Where $U_f(u,v)$, $\lambda$, $k$ and $z$ stand for E-field of final image, wavelength, wavenumber and the distance behind the lens in which the image is being formed respectively. Also, x,y and u,v are the spatial points in which the E-field is being calculated before and after the lens. Fig. 2 shows the details of the implementation structure and final derived images for each object.

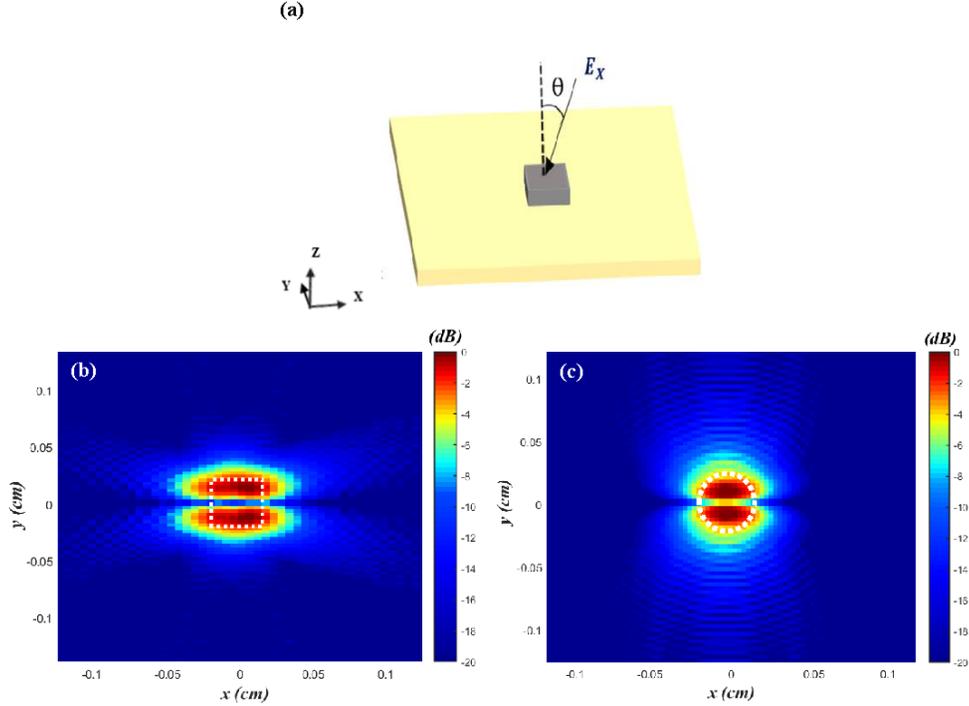

Fig. 2. (a) The implementation structure indicating the PEC object, incident plane wave $E_x$ and illumination angle $\theta$. (b),(c) Matlab code implementation results for PEC square and disk respectively : figures show the image formed at $z=35$ cm behind the lens, after E-field cx-polarized reflections of the PEC square object at $z=2$ m and PEC disk at $z=3$ m calculated using Eq. 1, went through the collimator lens.

As it can be seen in both Fig. 2. (b) and Fig. 2. (c), the cx-polarized reflection on the x-oriented edge that has the same direction as the polarization of the incident wave, is distinguishably higher than the cx-polarized reflection on y-oriented edge. In disk's results one can see the cx-polarized reflection's intensity becomes lower when getting far from its top and bottom points and close to its middles point at $y=0$ cm. Considering this results, it can be assumed that, illuminating the object by both polarizations simultaneously, will noticeably heighten the total cx-polarized reflections of the edges. This fact leads to a considerable opportunity to enhance the edge detecting process and consequently imaging accuracy.

### 3. SIMULATOIN RESULTS

To further evaluate the proposed idea, we have performed some near-field simulations by Altair, Feko using its Multilevel Fast Multipole Method (MLFM) and PO solver. A $3\times3\times0.5$ cm PEC object is placed on a $17\times17$ cm dielectric plane. Conductivity ($\sigma$), and relative permittivity ($\varepsilon r$) of the plane is chosen 39.42 and 5.6 respectively according to the conductivity and relative permittivity of the human skin at 100GHz, as represented in [22]. Also a 100 GHz plane wave simulates the source, illuminating the structure from x-axis. The illumination angle is chosen 40° relative to z-axis and the x-y plane at $z = 8$ mm is chosen as the observation plane of scattered field. The simulation structure is just the same as Fig. 2.

Since the incident plane wave is x-polarized, we have studied the x and y components of the scattered wave as co- polarized and cx-polarized reflections from the surface of the structure respectively. Fig. 3. (a) Shows the E-field co-polarized reflection of the object in absence of the body plane. It can be seen that the object is clearly apparent due to at least a 15dB difference between the object and free space. Next in Fig. 3. (b), we can see the E-field co-polarized reflection again, this time however, in the presence of the body plane. It is obvious that the object is less detectable when placed on human body. Fig.3. (c) and (d) represent the E-field cx-polarized reflection in the absence and presence of human plane that are almost the same with a near 2dB difference. This is because of the insignificancy of the cx-polarized reflection of human body as mentioned before. So it sounds reasonable to use cx-polarized reflection as a beneficial information in imaging process. In all figures, the object is remarked by a rectangular. Moreover, Fig. 3. (c) Indicates another remarkable point that is the x-oriented edges of the object have more significant reflection comparing to y-oriented ones as we anticipated according to PTD equations. Experimental results have been represented in next section for further demonstration.

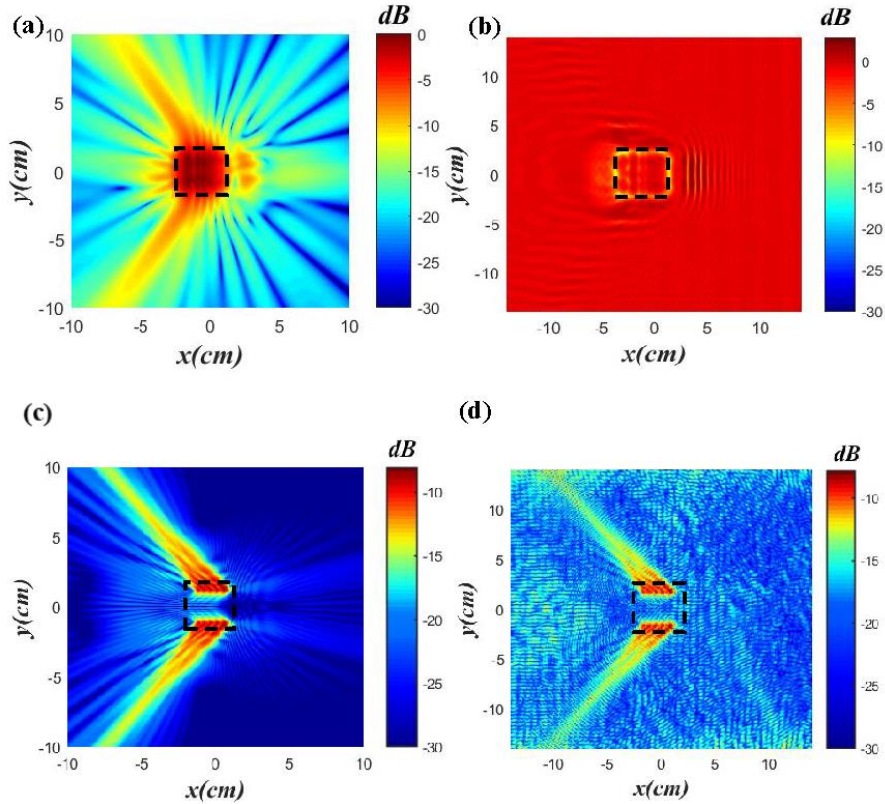

Fig. 3. (a) Co-polarized reflection of PEC object in absence of body plane. (b) Co-polarized reflection of PEC object in presence of human body. (c),(d) Cx-polarized of reflection of the object both in free space and placed on human plane. All figures show the E-field reflections at $z = 8$ mm.

## 4. EXPERIMENTAL RESULTS

In this section, first, we represents an overview of our imaging system. Next, we evaluate the efficacy of the polarimetry by doing dome experiments. The system shown in Fig. 4, benefits from IMPATT diode based sources with 0.8W output RF power provided with a horn antenna operating at 100GHz as a linear-polarized transmitter. After the antenna there is a Flat-top beam shaper (FBS) lens to convert the pseudo-Gaussian pattern of generated beam form the source to a uniform pattern in order to illuminate most part of the object from different angles [23]. This lens is made of PTFE $(n=1.45)$ with a 10cm diameter and performs collimation and beam-shaping at once. For achieving real-time detection, we use an array detecting system. This camera contains a $32\times32$ THz sensors fabricated on a $10\times10$ cm single wafer that mostly detects x-polarized waves. Each pixel consists of a GaAs high-mobility heterostructure that have a satisfying pixel-to-pixel deviation to each other, so their collected data will not be interfered. The camera detects the x-polarized. Again in front of the camera, there is an optical lens made of Teflon $(n=1.4)$ with a 30cm focal point, to collimate the beam received from the target. The effective field of view (FOV) of the system is an area of $70\times70$ cm on the target when the imaging distance is 3m.

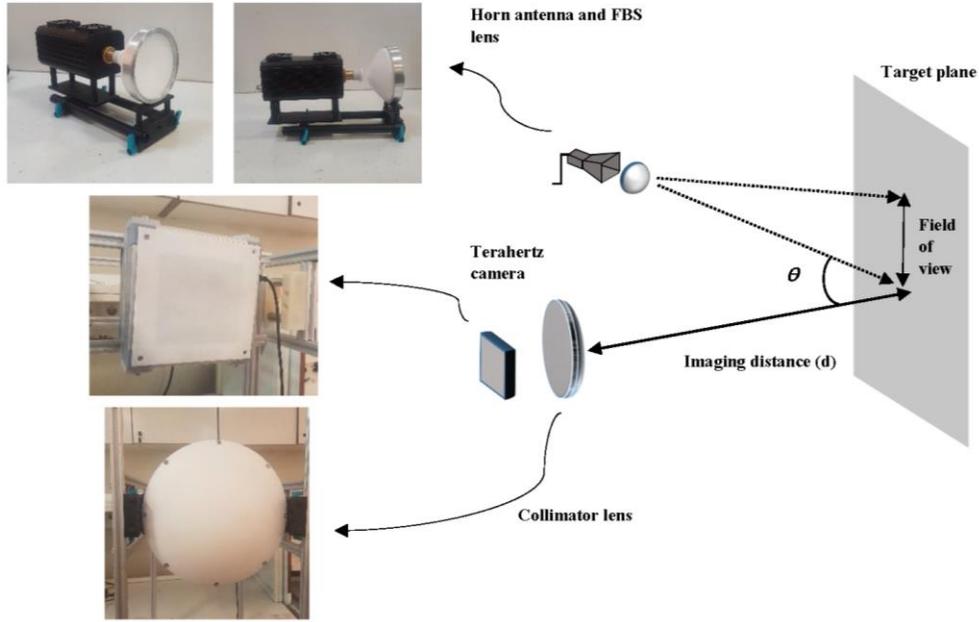

Fig. 4. Schematic of imaging system, indicating THz source, THz camera and collimator lens.

In order to demonstrate our proposed idea, we carried out some experiments that are represented in two different parts. In first, we used a $15\times12$ cm PEC rectangular and a 12cm diameter PEC disk as objects, placed on a polystyrene background and observed the cx and co-polarized reflections in two different tests. In each test, the detector is placed in front of the object and the sources is located next to the detector at the distance of 10 cm with a near-orthogonal ($\sim2°$) illumination angle relative to target plane's perpendicular axis. The imaging tests have been performed at the distance of 3m from the target with a 16 frame per second imaging frame rate. Fig. 5 shows the PEC object, co-polarized and cx-polarized imaging results respectively.

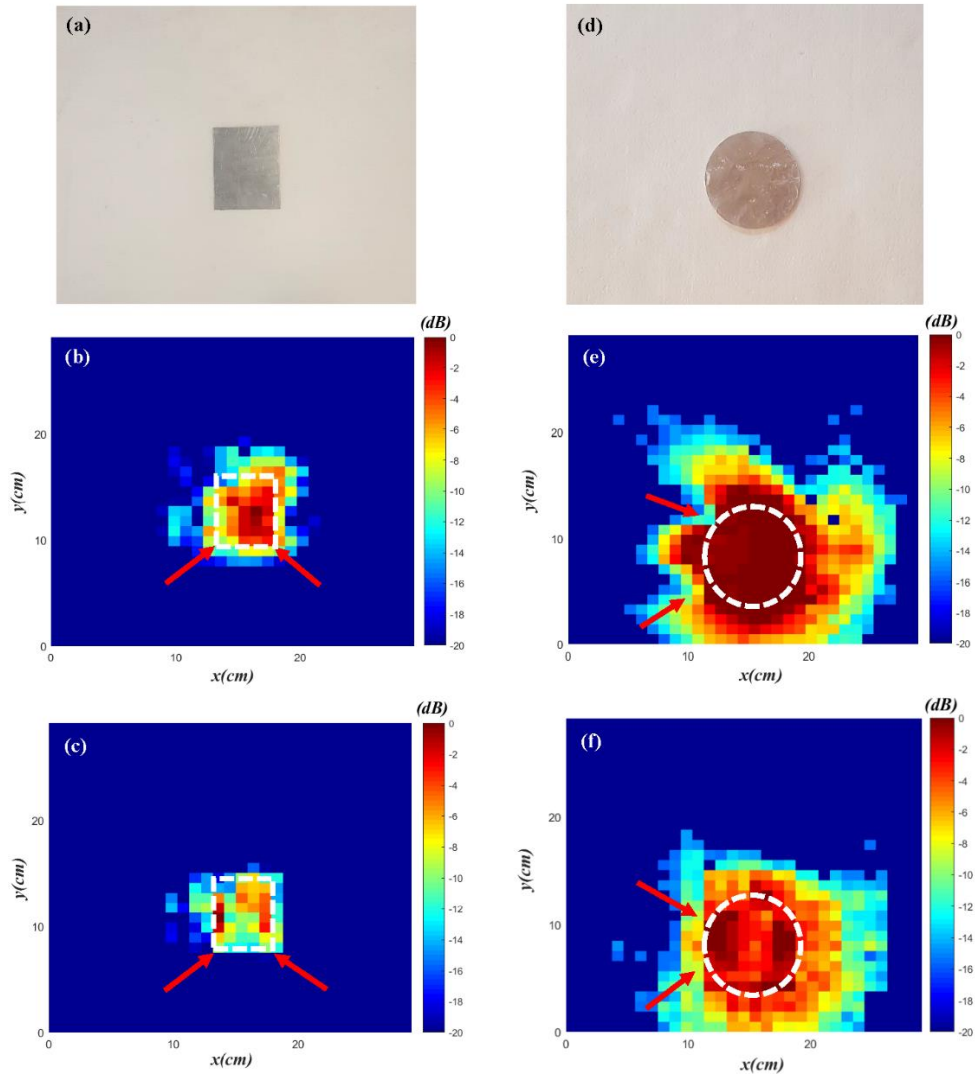

Fig. 5. (a),(b) PEC rectangular and disk respectively. (b), (c) show the co and cx-polarized reflections of rectangular object. (e) ,(f) show the co and cx-polarized reflections of disk object.

As it can be seen in Fig. 5. (b) And (c), in co-polarized image, the whole object, that is a PEC rectangular, has an almost equal reflection. However in cx-polarized image, the most significant reflection is due to object edges and the reflection intensity in middle of the object has declined. Fig. 5. (e) And (f) also show the same result about the disk object, confirming the proposed idea and simulation. Obviously, in this tests, the objects are detectable in both co-polarized and cx-polarized results, thus in next test, we investigated the effect of human body by placing a

7×7 cm object on a person and performing the same procedure. It should be mentioned that the polarization of the source is x-oriented in this test. Fig. 6 shows the imaging target, co-polarized and cx-polarized images respectively.

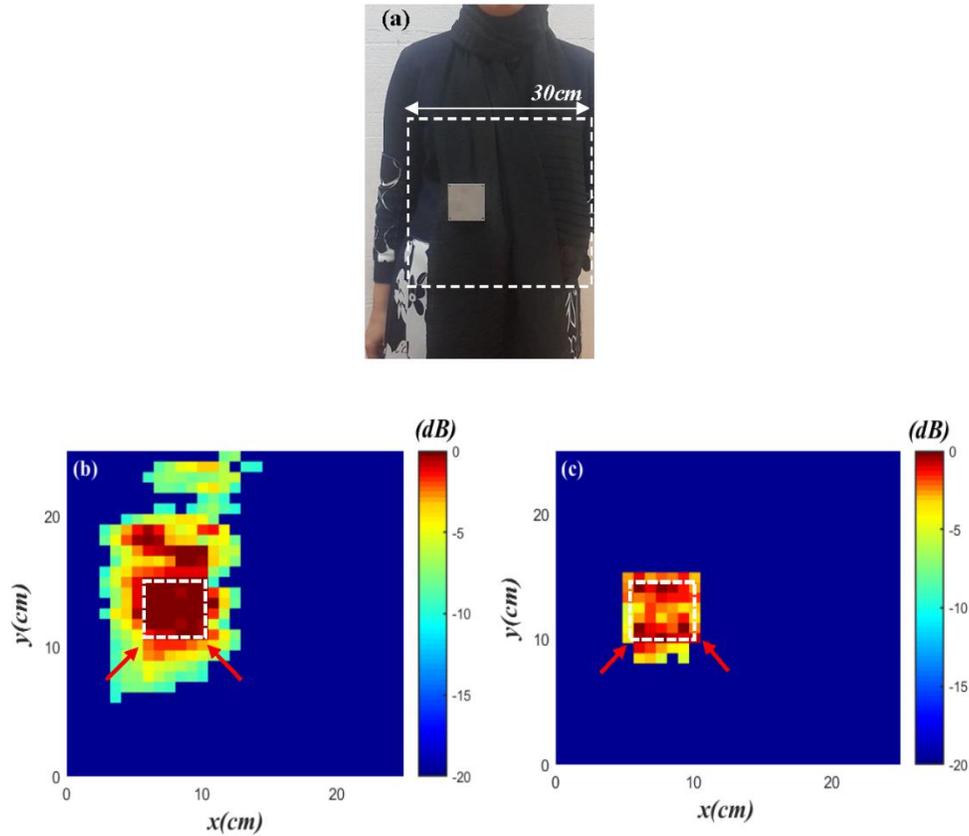

Fig. 6. (a) PEC object and Imaging area on the target (b) Co-polarized result containing both object and body reflection. (c) Cx-polarized reflection distinguishably indicating the object.

As it can be seen in co-polarized image, presence of the object is presumable but it's not clearly detectable due to reflections of the body. But in cx-polarized image, body reflections are eliminated and received data is exclusively related to the object. Such a difference can itself help distinguishing the object from the body but we took a step further to mostly benefit from this, as in next part of our experiment we illuminated the object with both x and y-polarized incident waves in order to have a good reflection from both horizontal and vertical edges. As it was mentioned before, since the horn antenna source generates an x-polarized linear wave, in order to accomplish our tests with a y-polarized incident wave, we changed its polarization to a y-polarized one by a 90° rotation relative to its central axis. Fig. 7 shows the results of both cx-polarized reflections from the object while the polarization of the source in first test is perpendicular to its polarization in the second one. A noticeable difference between vertical and horizontal edges can be observed as in the Fig. 7. (a), that the source is x-polarized, the horizontal edges have more intensity. In contrary, as we expected, Fig.7. (b) shows a more intensity on vertical edges due to the y-polarization of the source.

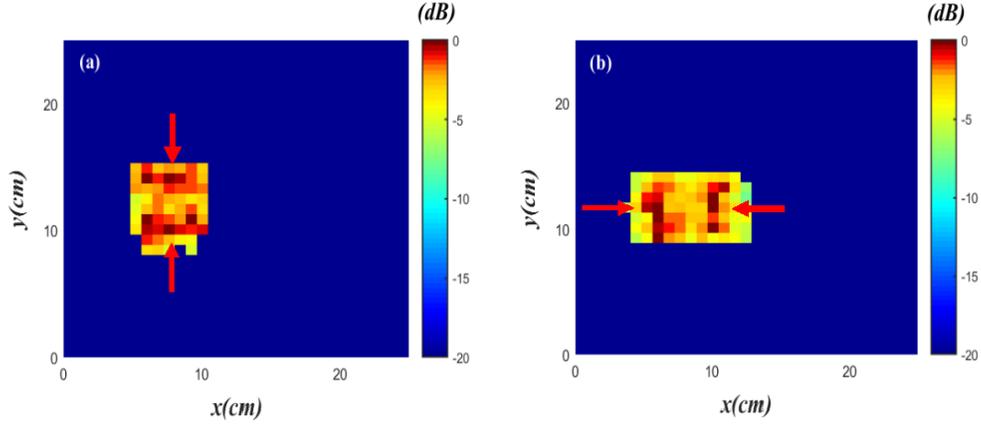

Fig. 7. (a), (b) Imaging result of cx-polarized reflection of the object illuminating by x and y-polarized source respectively.

## 5. IMAGE RECONSTRUCTION

To obtain the final image, we merged the objects' backscattered data of two tests, in which we used the source with different perpendicular polarizations in each test. The original imaging result is shown in Fig. 8(a). As we expected, the integrated image has a noticeable intensity on both edges indicating the approximate shape of the object.

Next we illustrated the superiority of using cx-polarized reflection by implementing an edge detection algorithm on the integrated image of Fig. 8(a). Among all edge detection algorithms we chose Canny algorithms due to its high quality performance [24]. Final results can be seen in Fig. 8(b).

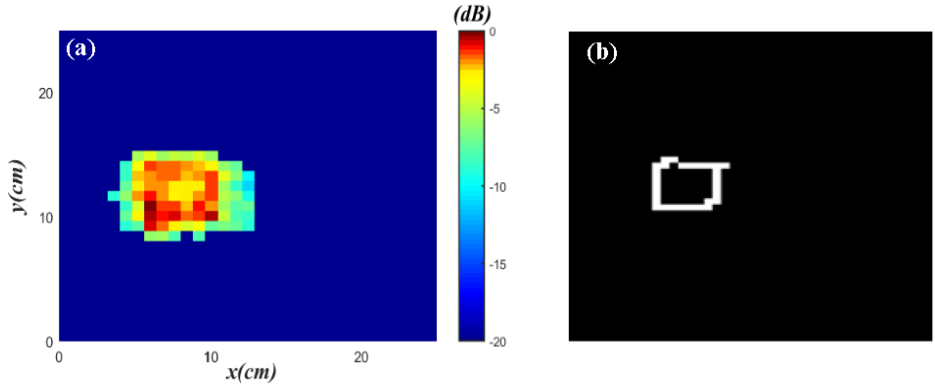

Fig. 8. (a) Integrated backscatter data collected from illuminating the target by both x and y-polarized incident wave (b) Result of implementing edge detection algorithm on co-polarized reflection of human body with the object

Fig. 8(b) clearly indicates that the object is almost correctly detected from cx-polarized reflection. However, in the result obtained from co-polarized reflection, object is not detectable from the body, demonstrating the significance of using cx-polarized reflection for a more accurate edge detection.

## 6. CONCLUSION

A Polarimetric imaging method is proposed in order to improve the accuracy of fast THz imaging. Relying on physical theory of diffraction, the idea of using the edge diffraction of PEC objects is proposed and both implemented by Matlab code and simulated using Multilevel Fast Multipole Method (MLFM) solver of Feko. Both implementation and simulation confirmed that the cx-polarized reflections of the edges of PEC objects can be beneficial in distinguishing them from human body. Some imaging tests are proposed to further illustrate the idea using IMPPAT diode sources provided with a horn antenna operating at 100 GHz. A collimator lens, a flat-top beam shaper (FBS) and a THz detector are other parts of our imaging system. Imaging results indicated a remarkably higher intensity of cx-polarized reflections from the edges of PEC objects making it discernable from human body. The effect of source's polarization on the edges' diffractions is also studied. Experimental results show an enhancement in cx-polarized reflections of the edges when the source illuminates the target with two orthogonal linear polarizations. Using slant linear waves or circular polarized waves in order to simultaneously illuminate all edges of the object can be used in further researches. The proposed method is promising for THz real-time imaging with high accuracy.